\documentclass[aps,prd,10pt,twocolumn,superscriptaddress,nofootinbib,letterpaper,longbibliography]{revtex4-2}

\usepackage{hyperref}
\usepackage{graphicx}
\usepackage{amsfonts,amsmath,amssymb,bm}
\usepackage{physics}
\usepackage{svg}
\usepackage{array}
\usepackage{dcolumn}
\usepackage{multirow}
\usepackage{booktabs}
\usepackage{lipsum}
\usepackage{float}
\usepackage{fontawesome5}
\usepackage{soul}
\newcolumntype{d}[1]{D{.}{.}{-1}}
\newcolumntype{C}[1]{>{\centering\let\newline\\\arraybackslash\hspace{0pt}}m{#1}}

\hypersetup{
  colorlinks  = true,
  urlcolor    = blue,
  linkcolor   = blue,
  citecolor   = red
}

\newcommand{\orcid}[1]
{\begingroup
\hypersetup{hidelinks}\href{https://orcid.org/#1}{\includegraphics[width=9pt]{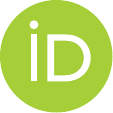}
} \endgroup}

\newcommand{\blue}[1]{\textcolor{black}{#1}}
\newcommand{\magenta}[1]{\textcolor{black}{#1}}
\newcommand{\sigmav}[0]{\langle \sigma v \rangle}
\newcommand{\B}[0]{$^8$B~}

\newcommand{\pderiv}[2]{\frac{\partial #1}{\partial #2}}
\newcommand{\bigparenthesis}[1]{\left(#1\right)}

\begin{document}

\title{A \texorpdfstring{$\nu$}{nu} look at the Sun: Probing the conditions of the solar core using \texorpdfstring{$^8$B}{8B} neutrinos}

\author{Melanie A. Zaidel \orcid{0009-0000-1925-7269}}
\email{zaidel.3@osu.edu}
\affiliation{Center for Cosmology and AstroParticle Physics (CCAPP), Ohio State University, Columbus, OH 43210}
\affiliation{Department of Physics, Ohio State University, Columbus, OH 43210}

\author{John F. Beacom \orcid{0000-0002-0005-2631}}
\email{beacom.7@osu.edu}
\affiliation{Center for Cosmology and AstroParticle Physics (CCAPP), Ohio State University, Columbus, OH 43210}
\affiliation{Department of Physics, Ohio State University, Columbus, OH 43210}
\affiliation{Department of Astronomy, Ohio State University, Columbus, OH 43210} 

\date{\textit{14 April 2025}}

\begin{abstract}
In the coming age of precision neutrino physics, neutrinos from the Sun become robust probes of the conditions of the solar core. Here, we focus on \B neutrinos, for which there are already high precision measurements by the Sudbury Neutrino Observatory and Super-Kamiokande. Using only basic physical principles and straightforward statistical tools, we estimate projected constraints on the temperature and density of the \B neutrino production zone compared to a reference solar model. We outline how to better understand the astrophysics of the solar interior using forthcoming neutrino data and solar models. \blue{Finally, we note that detailed forward modeling will be needed to develop the full potential of this approach.}
\end{abstract}

\maketitle

%%%%%%%%%%%%%%%%%%%%%%%%%%%%%%%%%%%%%%%%%%%%%%%%%%%%%%%%%%%%%%%%%%%%%%%
%%%%%%%%%%%%%%%%%%%%%%%%%%%%%%%%%%%%%%%%%%%%%%%%%%%%%%%%%%%%%%%%%%%%%%%

\section{Introduction}
\label{sec:intro}

The Sun is a foundational source in nuclear astrophysics~\cite{1958ApJ...127..551F, 1963ApJ...137..344B, Sears:1964zz, Bahcall:1989ks, 1988ccna.book.....R}. A longstanding program uses neutrino observations~\cite{Davis:1968cp, Kamiokande-II:1989hkh, Abazov:1991kq, GALLEX:1992jnx, Super-Kamiokande:1998qwk, SNO:2001kpb, SNO:2018fch, PandaX:2022aac, XENON:2024ijk} to directly probe the nuclear reactions that power the Sun's sustained luminosity~\cite{Bahcall:1963ohf, Bahcall:1964gx, Davis:1964hf}. Each of its nuclear-reaction chains produces a distinctive neutrino spectrum (depending upon laboratory nuclear physics) and flux (depending also upon the astrophysics of the solar interior)~\cite{1970ApJS...19..243C, 1987JPhB...20.6363S, Adelberger:1998qm, 2000ASIC..544..177D, Adelberger:2010qa, 2017Atoms...5...22P, Acharya:2024lke}. Success in this program is crucial to understanding all other main-sequence stars, for which we expect no direct observations of neutrinos and must therefore rely upon electromagnetic observations and theoretical modeling~\cite{1939isss.book.....C, 1964ApJ...139..306H,  1968pss..book.....C, 1983psen.book.....C, 1994sipp.book.....H, 2013sse..book.....K, Serenelli:2016dgz, Vinyoles:2016djt, 2021LRSP...18....2C}.

For several decades, this program was interrupted by the (exciting) diversion of the solar neutrino problem. In 1968, the first results from the Homestake radiochemical neutrino detection experiment showed a solar electron neutrino flux well below expectations from solar models~\cite{Davis:1968cp}. This tension persisted throughout efforts to better understand uncertainties in solar-model and nuclear physics, plus subsequent measurements of the neutrino flux\blue{es}. While proposed solutions to the solar neutrino problem included non-standard solar models, ultimately the correct answer was neutrino flavor mixing, which was first proposed for neutrinos in vacuum by Pontecorvo~\cite{Pontecorvo:1967fh} and later expanded to include matter effects by Mikheyev, Smirnov, and Wolfenstein (MSW)~\cite{Wolfenstein:1977ue, Mikheev:1986wj}. The definitive discovery of mixing was made by the Sudbury Neutrino Observatory (SNO), additionally confirming that the total \B neutrino flux agrees with predictions~\blue{\cite{SNO:2001kpb, SNO:2011hxd, Bellerive:2016byv}}. \magenta{Super-Kamiokande's (Super-K) measurements of \B neutrinos~\cite{Super-Kamiokande:2001ljr, Super-Kamiokande:2010tar, Super-Kamiokande:2023jbt} and Borexino's measurements of lower-energy neutrinos~\cite{Borexino:2000uvj, Borexino:2008dzn, Borexino:2011ufb, BOREXINO:2018ohr} also played decisive roles.}  Furthermore, the Kamiokande Liquid Scintillator Anti-Neutrino Detector (KamLAND) reactor experiment measured neutrino mixing parameters consistent with the large mixing angle oscillation (LMAO) solution for the MSW effect in the Sun\blue{~\cite{KamLAND:2002uet, KamLAND:2004mhv, KamLAND:2008dgz, KamLAND:2010fvi, KamLAND:2013rgu}}, up to some modest tension in the mass-squared splitting.

We will soon have an opportunity to return to the original program of probing the astrophysics of the solar core~\cite{Bahcall:1963ohf, Bahcall:1964gx, Davis:1964hf}.  Doing so may help resolve persistent discrepancies in solar helioseismology~\cite{Basu:2007fp, 2012ApJ...746...16V, Christensen-Dalsgaard:2018etv, 2025arXiv250406891B} and metallicity data~\cite{2014dapb.book..245B, 2014ApJ...787...13V, 2020arXiv200406365V}.  The Jiangmen Underground Neutrino Observatory (JUNO), starting operations in 2025, will improve uncertainties on some neutrino mixing parameters from 3--10\% to better than 1\%~\cite{JUNO:2015zny, JUNO:2024jaw}. Beyond that, new experiments, especially Hyper-Kamiokande~\cite{Hyper-Kamiokande:2016srs, 2018arXiv180504163H, Hyper-Kamiokande:2022smq} and the Deep Underground Neutrino Experiment (DUNE)~\cite{DUNE:2020mra, DUNE:2020lwj, DUNE:2020ypp, DUNE:2020txw, Capozzi:2018dat, JUNO:2022jkf, Meighen-Berger:2024xbx}, will greatly improve the precision of other neutrino mixing parameters. In the absence of indications of new physics, we therefore expect that the neutrino mixing parameters will be precisely known, independent of solar-neutrino data.  \textit{What, then, can solar-neutrino observations teach us about solar astrophysics?}

In this \blue{first} paper, we work towards model-independent probes of the astrophysics of the solar core based on neutrino data \blue{to estimate the potential of a more detailed study}, improving upon earlier work~\cite{1997PhDT.........9B, Balantekin:1997fr, Lopes:2013nfa, Lopes:2013sba, Laber-Smith:2022eih, Laber-Smith:2024hbc}. \blue{We aim to make the underlying physics clear and to provide motivations for further work, especially including with forthcoming experimental data. As we are estimating projected constraints, we make several approximations.} 

We focus on \B neutrinos, relying only on basic physical principles and straightforward statistics.  There are two basic ideas behind our work.  First, the total flux of \B neutrinos is very sensitive to the temperature of the solar core: $\phi \propto T_\textrm{c}^{\alpha}$, with $\alpha \approx 24$~\cite{Bahcall:1996vj}. Because SNO has determined the \blue{neutral current total} \B neutrino flux to better than 4\%~\cite{SNO:2011hxd}, this probes the solar core temperature to \magenta{0.2\% in the absence of other uncertainties. Taking into account uncertainties on laboratory inputs and solar modeling, the temperature uncertainty is more like 1\%~\cite{Haxton:2012wfz}, which can be improved in the future.} Second, because of how neutrinos mix, the electron neutrino survival probability at Earth depends on the solar density at production. For \B neutrinos, this effect becomes important at the low-energy end of the spectrum of scattered electrons in \magenta{Super-K}. Their most recent results --- using the full data period of Super-K-IV~\cite{Super-Kamiokande:2023jbt} --- thus allow for the best chance yet to probe the conditions of the solar core.

In Sec.~\ref{sec:calculations}, we calculate how the spectrum of scattered electrons in Super-K depends on the details of \B production, propagation, and detection. In Sec.~\ref{sec:probes}, we present our major result: the first combination of \blue{estimates of projected} constraints on solar structure in the plane of density and temperature, ($\rho, T$). While simple, this is new and provides important insights. We also outline broader studies of the solar core conditions. Finally, in Sec.~\ref{sec:conclusions}, we conclude and look towards the future of the solar astrophysics program.

%%%%%%%%%%%%%%%%%%%%%%%%%%%%%%%%%%%%%%%%%%%%%%%%%%%%%%%%%%%%%%%%%%%%%%%
%%%%%%%%%%%%%%%%%%%%%%%%%%%%%%%%%%%%%%%%%%%%%%%%%%%%%%%%%%%%%%%%%%%%%%%

\section{Calculation of the \texorpdfstring{$^8$B}{8B} signal}
\label{sec:calculations}

In this section, we present our calculations relating to \B neutrinos from the Sun. We begin with their birth in nuclear reactions, then follow their flavor mixing via the MSW effect, and finish with their detection in Super-K via neutrino-electron elastic scattering. The \href{https://github.com/melanieAzaidel/solarLMAO}{GitHub repository} provides short Jupyter Notebooks on \B neutrino production and detection, as well as the routines used to generate figures in this paper.

\begin{figure}[t]
    \centering
    \includegraphics[width=0.99\linewidth]{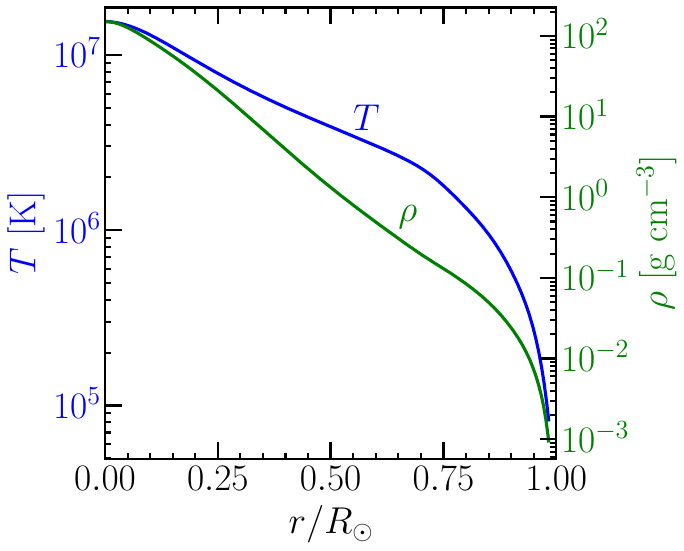}
    \caption{Density and temperature profiles from the reference solar model~\cite{Bahcall:2004pz}. Note different y-axes.}
    \label{fig:solarProfiles}
\end{figure}

As described in the introduction, \B neutrino observations at Earth depend on the astrophysics of the solar core. The total \B flux measured by SNO and the spectrum of recoil electrons in Super-K, respectively, allow us to probe the temperature and density in the region of \B production. Because of the limited neutrino data, we produce constraints relative to a reference solar model~\cite{Bahcall:2004pz}. Figure~\ref{fig:solarProfiles} shows its radial profiles of temperature and density.

%%%%%%%%%%%%%%%%%%%%%%%%%%%%%%%%%%%%%%%%%%%%%%%%%%%%%%%%%%%%%%%%%%%%%%%

\subsection{Production}
\label{subsec:production}

Solar \B neutrinos are produced via the following reactions of the pp-III chain:
\begin{align}
    ^7\text{Be} + \text{}^1\text{H} &\longrightarrow \text{} ^8\text{B} + \gamma 
    \label{eq:17_reaction} \\
    ^8\text{B} &\longrightarrow \text{} ^8\text{Be}^* + e^+ + \nu_\mathrm{e}.
    \label{eq:beta_decay}
\end{align}
While the neutrinos themselves are actually produced via Eq.~(\ref{eq:beta_decay}), the beta decay does not depend on the solar environment and proceeds much faster than the preceding fusion reaction. Consequently, the production of \B neutrinos is described by the rate of the fusion reaction in Eq.~(\ref{eq:17_reaction}), which depends on laboratory nuclear physics and the astrophysics of the solar core.  The \B neutrino spectrum is broad, with a peak near 6.5~MeV and an endpoint near 15~MeV~\cite{Winter:2004kf}.

%%%%%%%%%%%%%%%%%%%%%%%%%%%%%%%%%%%%%
%%%%%%%%%%%%%%%%%%%%%%%%%%%%%%%%%%%%%

The radial profile of the nuclear reaction rate density depends on the product of the number densities of the reactants, $n_i$, and the reaction rate factor, $\sigmav$:
\begin{equation}
    \Gamma_{ab}(r) = \frac{n_a n_b}{1 + \delta_{ab}}\sigmav.
    \label{eq:Gamma}
\end{equation}
\blue{We express the reaction rate density} in units of cm$^{-3}$~s$^{-1}$, where the Kronecker delta accounts for identical reactants. The number densities $n_i$ for various atomic species $i$ can be obtained by multiplying the mass density, $\rho$, by the mass fraction per species divided by the species mass, $X_i / m_i$. For $\sigmav$, we use a weighted average~\blue{\cite{1988ccna.book.....R}}, obtained by integration over the center-of-mass energy, $E$ of the product of the the fusion cross section, $\sigma(E)$, the velocity, $v(E)$, and the distribution of particles, $f(E)$, i.e.,
\begin{equation}
    \sigmav = \int_0^\infty dE \,
    \sigma(E) v(E) f(E).
    \label{eq:main_sigma_v}
\end{equation}
The energy-dependent behavior in the cross section is typically encapsulated via the experimentally determined astrophysical S-factor, defined by 
\begin{equation}
    \sigma(E) = \frac{S(E)}{E} \exp\bigparenthesis{-\sqrt{\frac{E_\mathrm{G}}{E}}},
    \label{eq:S_factor}
\end{equation}
where $E_G = 2m_r c^2 (\pi \alpha Z_a Z_b)^2$ is the Gamow energy (\blue{where the} fine structure constant \blue{is} $\alpha = e^2/\hbar c$, with the elementary charge $e$, reduced Planck constant $\hbar$, and speed of light $c$ \blue{are expressed in CGS units}), $m_r$ is the reduced mass of the pair of reactants, and $Z_i$ are their atomic numbers. The exponential function in Eq.~(\ref{eq:S_factor}) is often written as $\exp\bigparenthesis{-2\pi\eta}$ (the Gamow factor), where $\eta = Z_a Z_b (e^2/h) \sqrt{m_r c^2/(2E)}$ is the Sommerfeld parameter with the Planck constant $h$. For a non-relativistic Maxwellian distribution of particles, the reaction rate factor is
\begin{align}
\begin{split}
    \sigmav &= \frac{2\sqrt{2}}{\sqrt{m_r\pi}} \frac{1}{(k_B T)^{3/2}} \\
    &\times \int_0^\infty dE \, S(E) \exp\bigparenthesis{-\frac{E}{k_B T} - 2\pi \eta},
    \label{eq:sigma_v_full}
\end{split}
\end{align}
with the Boltzmann constant, $k_\mathrm{B}$.
%%%%%%%%%%%%%%%%%%%%%%%%%%%%%%%%%%%%%
\begin{figure}[t]
    \centering
    \includegraphics[width=0.99\linewidth]{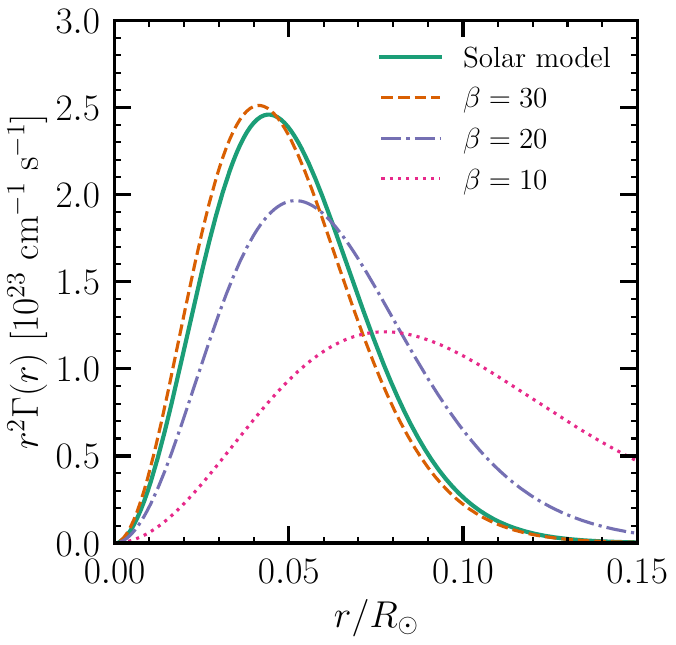}
    \caption{Full calculation of the \B neutrino production rate profile using Eqs.~(\ref{eq:Gamma}) and (\ref{eq:sigma_v_approximation}) compared to approximations that use only a simple scaling of the temperature profile. The normalizations of the latter curves are set equal to that of the full calculation.}
    \label{fig:NPZ}
\end{figure}
%%%%%%%%%%%%%%%%%%%%%%%%%%%%%%%%%%%%%

At astrophysical energies ($k_\mathrm{B} T \approx 1.3$~keV in the solar core), $S(E)$ is usually (in the absence of nuclear resonances) a slowly varying function. It is typical to write an effective S-factor, $S_\mathrm{eff}$, as Taylor series, for which coefficients are determined from theory and measurements. To describe reaction rates in the Sun, we use the compact approximation given in \blue{Ref.~\cite{1988ccna.book.....R}},
\begin{align}
\begin{split}
    \blue{\sigmav} &= \blue{\frac{4}{\sqrt{3}} \bigparenthesis{\frac{2\pi\alpha c Z_a Z_b}{m_r (k_\mathrm{B}T)^2}}^{1/3}S_\mathrm{eff}\exp(-\tau),}
    \label{eq:sigma_v_approximation_fundamental}
\end{split}
\end{align}
\blue{which can be numerically evaluated and reformulated~\cite{Bahcall:1989ks}} to produce 
\begin{align}
\begin{split}
    \sigmav &= 1.3005 \times 10^{-15} \left[\frac{Z_a Z_b}{A T_6^2}\right]^{1/3} \\ 
    &\times f_0 S_\mathrm{eff} \exp(-\tau) \text{ cm}^{3} \text{ s}^{-1}, 
    \label{eq:sigma_v_approximation}
\end{split}
\end{align}
where $A$ \blue{is $m_r$ expressed in atomic mass units} and $T_6 = T/(10^6\text{ K})$. The factor
\begin{equation}
    f_0 = \exp\bigparenthesis{0.188Z_a Z_b \varphi \rho^{1/2} T_6^{-3/2}}
\end{equation}
\blue{is included to} account for the screening of electrons, where $\rho$ is the mass density (in implicit units of g cm$^{-3}$) and $\varphi = \left[\sum_i (X_i Z_i^2 /A_i + X_i Z_i/A_i) \right]^{1/2}$. The latter quantity is usually labeled $\zeta$ in the literature, here we use $\varphi$ here to avoid confusion with a quantity connected with neutrino mixing (Sec.~\ref{subsec:propagation}). Next,
\begin{align}
\begin{split}
    S_\mathrm{eff} &= S_\mathrm{eff}(E_0) \approx S(0) \Biggl[1 + \frac{5}{12\tau} + \frac{S'}{S} \left(E_0 + \frac{35}{36}k_\mathrm{B}T\right)\\ &+ \frac{S'' E_0}{S} \left(\frac{E_0}{2} + \frac{89}{72}k_\mathrm{B}T \right) \Biggr]_{E = 0}
\end{split}
\end{align}
is the effective astrophysical S-factor \blue{expressed in keV~barns}, where $E_0 = 1.2204 \left(Z_1^2 Z_2^2 A T_6^2\right)^{1/3} \text{ keV}$ is the most probable energy of interaction (the Gamow peak), $\tau = 3E_0/(k_\mathrm{B}T) = 42.487 \left(Z_1^2 Z_2^2 A T_6^{-1}\right)^{1/3}$ is a dimensionless parameter convenient for the power series expansion of the S-factor, where $S(E)$, $S' \equiv dS/dE$, and $S'' \equiv d^2 S/dE^2$ are all evaluated at $E = 0$~\blue{\cite{Burbidge:1957vc}}. Recommendations for the S-factors can be found in reviews of solar fusion~\cite{Adelberger:1998qm, Adelberger:2010qa, Acharya:2024lke}. In the literature, the S-factor for the reaction in Eq.~(\ref{eq:17_reaction}) is referred to as $S_{17}$. We use the central value recommended in Ref.~\cite{Acharya:2024lke} and neglect its uncertainties (see Sec.~\ref{subsec:future})\blue{, anticipating further experimental improvements}. Finally, the factor $\exp(-\tau)$ arises from the Gamow peak.

Figure~\ref{fig:NPZ} shows our full calculation of the \B neutrino production rate profile (which includes the geometrical factor of $r^2$ but not the angular factor of $4\pi$), following Eqs.~(\ref{eq:Gamma}) and (\ref{eq:sigma_v_approximation}), where we take inputs from the reference solar model~\cite{Bahcall:2004pz}.  The \textit{neutrino production zone} is defined by the radii where this function is large.  We have checked that our calculation in the solid line closely matches the shape of the similar function in Ref.~\cite{Bahcall:2004pz} (not shown).

Figure~\ref{fig:NPZ} also shows an important point: that the full profile can be well approximated by simply $\blue{\Gamma(r) \propto T(r)^\beta}$. This may be somewhat surprising, as the calculations above show that the temperature and density enter the calculations in multiple ways.  To understand why this approximation works well, a few insights are needed.  First, we can largely ignore the density dependence that appears in Eq.~(\ref{eq:Gamma}). In the radial range 0--0.1$R_\odot$, the number density of $^1$H is only slightly falling and that of $^7$Be falls by about an order of magnitude.  While the latter sounds important, it can be accommodated by a moderate increase in $\beta$, as shown in the figure.  We find that the solar model curve obtained from using Eqs.~(\ref{eq:Gamma}) and (\ref{eq:sigma_v_approximation}) can be largely recovered by using $\beta \approx 27$.

Second, the reaction rate factor, $\sigmav$, as specified in Eq.~(\ref{eq:sigma_v_approximation}), can be taken to be independent of density and to depend on temperature primarily through the terms $\sigmav \propto T(r)^{-2/3} \exp(-\tau)$.  The variations in $f_0$ and $S_\mathrm{eff}$ with temperature and density are comparatively very weak.  Then we can write
\begin{equation}
    \sigmav = \sigmav_0 \bigparenthesis{\frac{T(r)}{T_0}}^\beta,\text{ } \beta = \pderiv{\ln \sigmav}{\ln T(r)}
    \label{eq:powerLaw}
\end{equation}
where $\sigmav_0$ is a normalization factor and $T_0$ is some reference temperature~\cite{Kippenhahn:2012qhp}. Following taking the log of Eq.~(\ref{eq:powerLaw}) and then taking the log derivative \blue{with respect to} temperature, it can be shown that
\begin{equation}
    \beta \approx \frac{\tau}{3}.
\end{equation}
For the fusion of $^7$Be and a proton near the solar center, $\tau \approx 41$, so $\beta \approx 14$ is the expected value assuming only \blue{the} temperature dependence. (We have checked that including the radial composition profiles for $^1$H and especially $^7$Be leads to close agreement with the full calculation.)

We must distinguish between two temperature exponents and how they are used.  Above, $\beta$ is applied to the \textit{local temperature}, $T(r)$.  Separately, the total \B neutrino flux scales as $T_\textrm{c}^\alpha$, where $T_\textrm{c}$ is the \textit{central temperature}~\cite{Bahcall:1987jc}. Across many contemporary solar models, it is found that $\alpha \approx 24 \pm 5$~\cite{Bahcall:1996vj} (at the time of Ref.~\cite{Bahcall:1989ks}, it was said that $\alpha \approx 18$, which may be a more familiar value).

In lieu of developing our own self-consistent non-standard solar model, we exploit the fact that $\Gamma(r) \propto T(r)^\beta$ to develop a semi-model-independent approach to probe the densities of the \B production region.  While a curve with $\beta \approx 14$ in Fig.~\ref{fig:NPZ} would be shifted to slightly larger radii than the full calculation, the width of the distribution is approximately correct. Therefore, as a first step towards a more sophisticated approach that will be possible with future data, we will assume that the \textit{uncertainties} we derive with our approximate approach will be informative, even if the \textit{values} are somewhat shifted.

As we show in Sec.~\ref{subsec:propagation}, varying $\beta$ changes the electron neutrino survival probability at Earth because it changes the density at production.  To \blue{calculate the estimates of this paper}, we neglect the fact that varying $\beta$ would also change the total neutrino flux at Earth because it changes the temperature at production; instead, we keep the total flux at the central value measured by SNO. As discussed in Sec.~\ref{subsec:future}, in future work \blue{the effect of $\beta$ on the flux} can be taken advantage of. 

%%%%%%%%%%%%%%%%%%%%%%%%%%%%%%%%%%%%%%%%%%%%%%%%%%%%%%%%%%%%%%%%%%%%%%%

\subsection{Propagation}
\label{subsec:propagation}

Neutrinos born in the electron-rich environment of the solar core experience flavor mixing due to the MSW effect. This mixing depends on three particle parameters, where we use the values taken from reactor experiments: \blue{$\tan^2(2\theta_{12}) = 0.436^{+0.029}_{-0.025}$, $\Delta m^2_{21} = (7.53 \pm 0.18) \times 10^{-5}$~eV$^2$, and $\sin^2(\theta_{13}) = 0.022 \pm 0.001$~\cite{Decowski:2016axc, Esteban:2024eli}}. We quote the uncertainties to indicate their impressive precision, but neglect them in our calculation because we focus on the near-future situation in which they will be much smaller, following measurements by JUNO and other experiments.  We use the mixing parameters from reactor experiments because they are independent of solar astrophysics; our results do not change much if we instead use the mixing parameters from solar experiments.  \blue{Further, we assume that no new-physics effects (e.g., Refs.~\cite{Friedland:2004pp, Maltoni:2015kca, Gann:2021ndb}) change the neutrino survival probability.}

The first discussions of probing the solar density using neutrino mixing were in Refs.~\cite{1997PhDT.........9B, Balantekin:1997fr}, which considered the limits of large and small mixing angles, respectively. The physics of this problem for the LMAO case was developed further in Refs.~\cite{Lopes:2013nfa, Lopes:2013sba}, which found best-fit densities in some tension with solar models.  More recently, Refs.~\cite{Laber-Smith:2022eih, Laber-Smith:2024hbc} developed new statistical tools for this problem, finding overall compatibility with solar models.  Our focus is on developing a more detailed approach with clearer physical and statistical insights, plus we also develop the first joint probes for the $\rho, T$ plane (Sec.~\ref{sec:probes}).

\begin{figure}[t]
    \centering
    \includegraphics[width=0.99\linewidth]{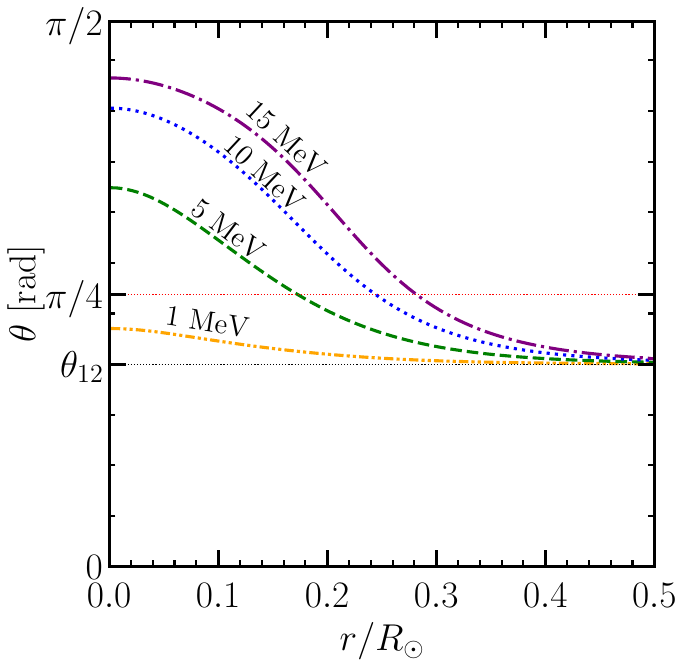}
    \caption{Variation of the matter angle throughout the Sun for three representative neutrino energies for \B and one much lower energy for contrast. For a given neutrino energy, the MSW resonance occurs when the matter angle is $\pi/4$.  That criterion and the vacuum mixing angle are noted with horizontal dotted lines.}
    \label{fig:matterAngle}
\end{figure}

Figure~\ref{fig:matterAngle} shows how the mixing angle is modified in matter due to the MSW effect, the core idea of th\blue{e}se studies.  The matter angle $\theta$ depends on the solar electron number density as~\cite{Balantekin:1996ag, Giunti:2007ry}
\begin{equation}
    \cos{(2\theta(E_\nu,n_\mathrm{e}))} = \frac{-\magenta{\psi}}{\sqrt{\Lambda^2 + \magenta{\psi}^2}}
\end{equation}
where $\Lambda = \sin{(2\theta_{12})}$, $\magenta{\psi} = \zeta(E_\nu,n_\mathrm{e}) - \cos{(2\theta_{12})}$, and $\zeta = 2\sqrt{2}G_F E_\nu n_\mathrm{e}/\Delta m^2_{21}$, with $E_\nu$ being the neutrino energy and $G_F$ being the Fermi coupling constant, and $n_\mathrm{e}$ is the number density of electrons. The electron number density profile, $n_\mathrm{e}(r)$, is related to the mass density, $\rho(r)$, via the mean molecular weight per electron, $\mu_\mathrm{e}(r)$, \blue{which} encodes the composition profiles from the reference solar model. 

In vacuum, the matter mixing angle simply becomes the standard mixing angle, $\theta_{12}$. Note that \B neutrino production occurs at small radii (Fig.~\ref{fig:NPZ}), while the MSW resonances occur at moderate radii (Fig.~\ref{fig:matterAngle}). This turns out to be critical, as detailed below.

The three-flavor electron neutrino survival probability is approximately~\cite{Bilenky:2010zza, KamLAND:2010fvi, Maltoni:2015kca}
\begin{equation}
    P_{\nu_\mathrm{e}}(E_\nu) = \cos^4{(\theta_{13})} P_{2\nu_\mathrm{e}}(E_\nu) + \sin^4{(\theta_{13})},
   \label{eq:threeFlavor}
\end{equation}
where we have neglected a small effect due to solar matter effects on $\theta_{13}$, and where the two-flavor component is~\cite{Parke:1986jy}
\begin{equation}
    P_{2\nu_\mathrm{e}}(E_\nu) = \frac{1}{2}\left[ 1 + \langle \cos{(2\theta(E_\nu))} \rangle_\mathrm{src} \cos{(2\theta_{12})}  \right].
    \label{eq:twoFlavor}
\end{equation}
The two factors in Eq.~(\ref{eq:twoFlavor}) reflect the initial and final matter angles, where the initial one must be averaged over the neutrino production rate profile and the final one is the same as in laboratory experiments. (We neglect neutrino regeneration effects in Earth, which would increase the survival probability by a few percent in this energy range~\cite{Goswami:2004cn, Maltoni:2015kca}, \blue{and consequently only use daytime observations.}) Importantly, Eq.~(\ref{eq:threeFlavor}) depends only on the (averaged) matter density at \B production, and not the densities along the neutrino trajectories.

%%%%%%%%%%%%%%%%%%%%%%%%%%%%%%%%%%%%%
%%%%%%%%%%%%%%%%%%%%%%%%%%%%%%%%%%%%%

Before discussing the validity and averaging of Eqs.~(\ref{eq:threeFlavor}) and  (\ref{eq:twoFlavor}), we note some key physics points, building on Ref.~\cite{1997PhDT.........9B}.  The only place that the solar density appears is in the initial matter angle, $\cos{(2\theta(E_\nu,n_\mathrm{e}))}$; this is also the only place that the neutrino energy appears.  Consider three cases \textit{where observations of neutrino mixing would reveal nothing about the solar interior}:
\begin{itemize}

\item If $\Delta m^2_{21}$ had been much larger (equivalent to the solar core densities or typical neutrino energies being much smaller --- see the 1~MeV curve in Fig.~\ref{fig:matterAngle} --- or to having the MSW resonance density be much larger than the \B production density), then $\cos{(2\theta(E_\nu,n_\mathrm{e}))} \rightarrow \cos{(2\theta_{12})}$ and $P_{2\nu_\mathrm{e}}$ is a constant ($= \frac{1}{2}[1 + \cos^2({2\theta_{12}})]$).

\item If $\Delta m^2_{21}$ had been much smaller (equivalent to the opposite conditions), then $\cos{(2\theta(E_\nu,n_\mathrm{e}))} \rightarrow -1$ and $P_{2\nu_\mathrm{e}}$ is a constant ($= \sin^2({\theta_{12}})$).

\item If $\Delta m^2_{21}$ had been somewhat smaller and fine-tuned so that the MSW resonance density was equal to the \B production density, then $\cos{(2\theta(E_\nu,n_\mathrm{e}))} \rightarrow 0$ and $P_{2\nu_\mathrm{e}}$ is a constant ($= 1/2$). There is a slight exception in this case, because some density dependence could arise through source averaging.

\end{itemize}
It is therefore fortunate that we are in the situation where \B neutrino mixing via $\Delta m^2_{21}$ does probe the solar interior.  Incidentally, the first bullet above is part of why mixing via the much larger $\Delta m^2_{31}$ is much less important; also, $\sin^2(\theta_{13})$ is small.

%%%%%%%%%%%%%%%%%%%%%%%%%%%%%%%%%%%%%
\begin{figure}[t]
    \centering
    \includegraphics[width=0.99\linewidth]{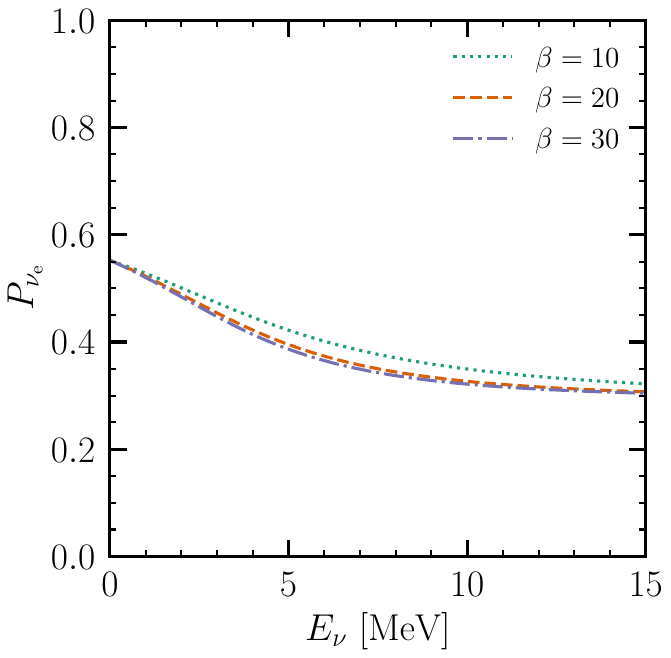}
    \caption{Electron neutrino survival probability at Earth as a function of neutrino energy for representative parameterizations of the neutrino production zone.}
    \label{fig:survivalProbability}
\end{figure}
%%%%%%%%%%%%%%%%%%%%%%%%%%%%%%%%%%%%%

The form of the survival probability in Eqs.~(\ref{eq:threeFlavor}) and  (\ref{eq:twoFlavor}) is valid under two conditions.  First, the neutrino propagation must be adiabatic, meaning that the probability to jump from one mass eigenstate to another --- which depends on the density \textit{along} the neutrino trajectory --- must be negligible, as it is for the LMAO mixing parameters~\cite{deHolanda:2004fd, Giunti:2007ry}.  (For small mixing angles, in contrast, the survival probability can depend only on the density --- more accurately, the variation of the density --- along the neutrino trajectory~\cite{1997PhDT.........9B, Balantekin:1997fr}.) Second, we must be able to neglect oscillatory terms in the neutrino survival probability~\cite{Balantekin:1996ag}.  As part of this, we note that the neutrino oscillation length in matter,
\begin{equation}
    L_\mathrm{M}^{\mathrm{osc}} = \frac{4\pi E}{\Delta m_\mathrm{M}^2}, \label{eq:oscillationLength}
\end{equation}
should be negligible compared to the size of the neutrino production zone. The effective mass splitting is
\begin{equation}
    \Delta m_\mathrm{M}^2 = \Delta m_{21}^2\sqrt{(\cos{(2\theta_{12})} - \zeta)^2 + \Lambda^2}.
\end{equation}
The size of the neutrino production zone is of order $10^{-1}R_\odot$, while the oscillation length in matter of the most energetic solar neutrinos does not exceed $10^{-3} R_\odot$.

The initial-state cosine must be averaged over the \B neutrino production zone as
\begin{equation}
    \langle \cos{2\theta(E_\nu, \beta)} \rangle_\mathrm{src} = \int_{r_\mathrm{min}}^{r_\mathrm{max}} dr \, r^2 T(r)^\beta  \cos\left(2\theta(E_\nu, n_\mathrm{e})\right).
    \label{eq:source_term}
\end{equation}
The neutrino production rate profile, which is proportional to $r^2 T(r)^\beta$, is normalized to unity for use in the above integral (the normalization is taken into account in Sec.~\ref{subsec:detection}).  As shown above, it is a reasonable approximation to characterize the neutrino production rate profile in this way, neglecting smaller effects due to the radial variation of density and composition.

Figure~\ref{fig:survivalProbability} shows the electron neutrino survival probability as a function of neutrino energy for representative neutrino production zones. For the electron number density profile from the reference solar model, the largest differences in survival probability for different neutrino production zones occur in the intermediate energy range, around 3--8~MeV. Even so, it is difficult to distinguish between the deepest production zones, e.g., the highest values of $\beta$. This is due to the flattening of the $n_\mathrm{e}$ profile in the core.  That means that even though the $\beta = 30$ and $\beta = 20$ cases look different in Fig.~\ref{fig:NPZ}, they are hard to distinguish with \B neutrinos.

We close with another important point.  In Eqs.~(\ref{eq:threeFlavor}) and (\ref{eq:twoFlavor}), the solar density dependence is tied to the neutrino energy dependence.  However, this does not mean that measuring the energy dependence of the survival probability would allow us to invert for the radial density profile.  At each neutrino energy, a range of radii in the neutrino production zone contribute, but only the average survival probability can be probed. If the \B neutrino production zone were narrower in radius, the differences shown in Fig.~\ref{fig:survivalProbability} would be larger.  Even so, appreciable differences remain.

%%%%%%%%%%%%%%%%%%%%%%%%%%%%%%%%%%%%%%%%%%%%%%%%%%%%%%%%%%%%%%%%%%%%%%%

\subsection{Detection}
\label{subsec:detection}

For \B neutrinos, we focus on observations with Super-K, due to their high statistics and broad spectrum coverage~\cite{Super-Kamiokande:2023jbt}.  There are also observations with SNO~\cite{SNO:2005oxr}, \blue{KamLAND~\cite{KamLAND:2011fld}}, and \blue{Borexino~\cite{Borexino:2008fkj, Borexino:2017uhp, Gonzalez-Garcia:2023kva}}. Super-K, located deep underground in Japan, is a water-Cherenkov neutrino detector sensitive to MeV-scale neutrinos.  It has a fiducial volume of 22.5~kton.  Neutrinos from the Sun elastically scatter with electrons in the water via both charged- and neutral-current interactions.  The spectrum of recoil electrons can be predicted using the neutrino production and propagation calculations above, the physics of weak interactions, and detector effects. 

It is important to note that Super-K (and Borexino) do not directly measure the electron neutrino survival probability.  (Below, we discuss the potential of using SNO's charged-current neutrino-deuteron data.)  First, because the differential cross section for neutrino-electron scattering is \blue{mostly flat in electron energy}, measurements can be made only as a function of electron energy, not neutrino energy.  Second, as discussed below, the non-electron-neutrino flavors contribute at a small level to the observed data.  Third, in order to normalize expectations in a semi-model-independent way, they rely on SNO's measurement of the total flux via neutral-current neutrino-deuteron interactions; \textit{this is only available for $^\textit{8}B$ neutrinos}.  Finally, just as a point of presentation, Super-K typically shows the ratio of the observed recoil electron spectrum to its theoretical expectation (relatively flat), rather than the recoil electron spectrum itself (relatively steep).

We begin with the initial electron spectrum,
\begin{equation}
    R(\mathcal{T}) = \int_{\mathrm{E}_\nu^\mathrm{min}}^{\mathrm{E}_\nu^\mathrm{max}} d\mathrm{E}_\nu \ \Phi_\nu (\mathrm{E}_\nu) \mathcal{P}(\mathcal{T},\mathrm{E}_\nu),
    \label{eq:theory_spectrum}
\end{equation}
where $\mathcal{T}$ is the true (before energy resolution effects) electron kinetic energy, $\mathrm{E}_\nu^\mathrm{min} = \frac{1}{2} \left(\mathcal{T} + \sqrt{\mathcal{T}^2 + 2m_\mathrm{e}\mathcal{T}} \right)$ is given by kinematics with the electron mass $m_\mathrm{e}$, and $\mathrm{E}_\nu^\mathrm{max}$ is the cutoff of the neutrino energy spectrum. The differential solar neutrino flux,
\begin{equation}
    \Phi_\nu (\mathrm{E}_\nu) = \phi_\mathrm{tot} \mathrm{f}(\mathrm{E}_\nu),
\end{equation}
is defined by $\phi_\mathrm{tot} = (5.25 \pm 0.20) \times 10^6 \text{ cm}^{-2} \text{ s}^{-1}$ which is the total \B neutrino flux measured by SNO~\cite{SNO:2011hxd}, and $\mathrm{f}(\mathrm{E}_\nu)$ which is the \B neutrino energy spectrum~\cite{Winter:2004kf}.

Neutrino oscillations and elastic scattering are accounted for via 
\begin{align}
\begin{split}
    \mathcal{P}(\mathcal{T},\mathrm{E}_\nu) &= \biggl[P_{\nu_\mathrm{e}}(\mathrm{E}_\nu) \frac{d\sigma_e}{d\mathcal{T}} (\mathcal{T},\mathrm{E}_\nu)\\ &+ \left(1 - P_{\nu_\mathrm{e}}(\mathrm{E}_\nu)\right)\frac{d\sigma_\mu}{d\mathcal{T}} (\mathcal{T},\mathrm{E}_\nu) \biggr],
\end{split}
\end{align}
where
$P_{\nu_\mathrm{e}}(\mathrm{E}_\nu)$ is the survival probability of electron neutrinos as a function of neutrino energy as given in Eq.~(\ref{eq:threeFlavor}) and $d\sigma_\chi/d\mathcal{T} (\mathcal{T},\mathrm{E}_\nu)$ is the differential cross section~\cite{Bahcall:1995mm}. Here, $\chi$ is $\mathrm{e}$ or $\mu/\tau$ for the electron and muon/tau flavors. Because the latter two flavors do not have charged-current interactions with electrons, their differential cross section is $\approx$6 times lower than for electron neutrinos.

To account for energy resolution effects in Super-K, Eq.~(\ref{eq:theory_spectrum}) is passed into
\begin{equation}
    S(\mathcal{T}) = \int_{{T_{\mathrm{e,min}}}}^{{T_\mathrm{e,max}}} dT_\mathrm{e} \ G(T_\mathrm{e},\mathcal{T},\sigma(\mathcal{T})) R(\mathcal{T}),
    \label{eq:resolvedSpectrum}
\end{equation}
where $T_\mathrm{e}$ is the measured kinetic energy. The energy resolution is taken into account via Gaussian smearing~\cite{Bahcall:1996ha},
\begin{align}
\begin{split}
    G(T_\mathrm{e},\mathcal{T},\sigma(\mathcal{T})) &= \frac{1}{\sigma(\mathcal{T})\sqrt{2\pi}} \\ &\times \exp\left[-\frac{(\mathcal{T} - T_\mathrm{e})^2}{2(\sigma(\mathcal{T}))^2}\right],
\end{split}
\end{align}
where 
\begin{align}
\begin{split}
    \sigma(\mathcal{T}) &= -0.5525 + 0.3162\sqrt{\mathcal{T}+ \mathrm{m}_\mathrm{e}} \\ &+ 0.04572(\mathcal{T} + m_\mathrm{e})
\end{split}
\end{align}
models the energy resolution during the full Super-K-IV period~\cite{Super-Kamiokande:2023jbt}. The limits of integration in Eq.~(\ref{eq:resolvedSpectrum}) are chosen to eliminate energy-resolution-induced smoothing effects on the edges of the spectrum. 

Finally, the spectrum is discretized according to Super-K's data bins as $\left(\Delta N/\Delta T \right)_i$, the event rate in the $i^\text{th}$ bin, in units of event day$^{-1}$ (22.5 kton)$^{-1}$ MeV$^{-1}$. We compute the final spectrum using 
\begin{equation}
    \left(\frac{\Delta N}{\Delta T} \right)_i = \frac{\mathrm{N}_\mathrm{e}}{\mathrm{M}_\mathrm{det}\Delta T_i} \int_{T_i^\mathrm{left}+\delta}^{T_i^\mathrm{left} + \Delta T_i +\delta} d\mathcal{T} \ S(\mathcal{T}),
    \label{eq:ourModel}
\end{equation}
where $\Delta T_i$ is the width of the $i^\mathrm{th}$ Super-K kinetic energy bin, $T_i^{\mathrm{left}}$ is the left edge of the bin, and $\delta$ is the uncertainty in the absolute energy scale of the detector~\cite{Bahcall:1999ed}, which we set to zero for simplicity \blue{to compute our initial estimates, because variations in $\delta$ negligibly affect our results}. The prefactors outside the integral include $\mathrm{N}_\mathrm{e} = 7.521 \times 10^{33}$ as the number of electrons and $\mathrm{M}_\mathrm{det} = 22.5$ kton as the total mass of water in Super-K's fiducial volume.

In our calculations, we consider only the signal from \B neutrinos. Those from the \textit{hep} reaction are produced somewhat further out in the Sun ($\approx$$0.15 R_\odot$)~\cite{Gann:2021ndb} and their spectrum extends to higher neutrino energies ($E_\nu^{\mathrm{max}} \approx 19$~MeV for $hep$~\cite{BahcallHepSpectrum} compared to $E_\nu^{\mathrm{max}} \approx 15$~MeV for $^8$B). While \textit{hep} neutrinos may be helpful in the future to study the solar interior, \magenta{their contribution to the Super-K and SNO data is not yet detected~\cite{Super-Kamiokande:2023jbt, SNO:2020gqd}}.  We thus exclude the last three Super-K bins, covering 13.49--19.49~MeV.  Additionally, we consider only statistical uncertainties on the Super-K observed recoil electron spectrum. Energy-correlated systematic uncertainties become significant only in the highest-energy bins~\cite{Super-Kamiokande:2023jbt}, which we have cut from our analysis. Energy-uncorrelated systematics are comparable to statistical uncertainties below $\approx$8~MeV.

\blue{In this first paper, we focus on calculating estimates of the temperature and density in the \B neutrino production zone. A more careful uncertainty analysis for the predicted \B signal is the subject of future work, so we only comment on some of our approximations. For neutrino production, we neglect how changing the site of neutrino production affects the implied total neutrino flux. We made this choice to keep the flux and survival probability observables separate to better illustrate the underlying physics. In principle, it would be interesting to consider how different \B production rate profiles would change the active-flavor flux to produce elliptical confidence regions on the temperature-density plane in the solar interior. For neutrino propagation, we are only interested in probing the density of the solar interior, and therefore do not consider Earth-effects, which would negligibly impact our final fit on $\beta$. For neutrino detection, our choice to exclude systematic uncertainties affects the final fit of $\beta$ well within $1\sigma$ limits.}

We compare the results of our simple treatment to the Super-K collaboration's careful Monte Carlo model of their detector (in the absence of neutrino mixing). For this case, the choice of $\beta$ does not enter ($P_{\nu_\mathrm{e}}(\mathrm{E}_\nu) = 1$). We first take the ``day" observed rate from Table IX of Ref.~\cite{Super-Kamiokande:2023jbt}. We compare our numerical evaluation in Eq.~(\ref{eq:ourModel}) to the expected event rate in Super-K due to \B neutrinos from the same table, and divide the observed rate by each model. Our calculation (detailed in the GitHub repository) is consistent with the Super-K model within their statistical uncertainties.

\begin{figure}[t]
    \centering
    \includegraphics[width=0.99\linewidth]{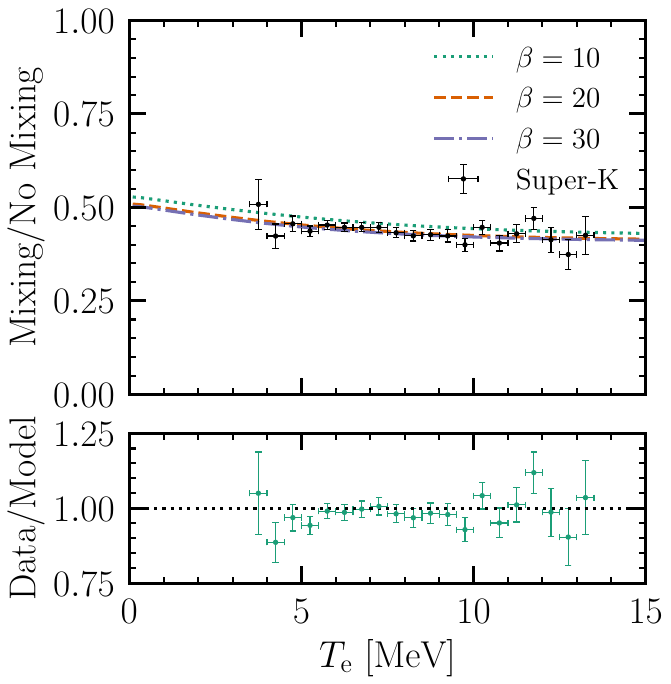}
    \caption{\textbf{Top panel}: Predictions of the electron spectrum for different $\beta$ values, calculated by taking the ratio of Eq.~(\ref{eq:resolvedSpectrum}) with and without neutrino flavor mixing (see text). \textbf{Bottom panel}: Ratio between the Super-K observations~\cite{Super-Kamiokande:2023jbt} of the recoil electron spectrum and our prediction from Eq.~(\ref{eq:ourModel}) with neutrino flavor mixing, using $\beta = 15$ as a representative case.}
    \label{fig:predictedSpectrum}
\end{figure}

Figure~\ref{fig:predictedSpectrum} (top panel) shows our predictions for the spectrum of scattered electrons from Eq.~(\ref{eq:resolvedSpectrum}) (with and without neutrino flavor mixing), compared to Super-K observations. The data points are a ratio between the observed spectrum in Super-K (with mixing implied) and their model (no mixing). Error bars show statistical uncertainties only.  The upturn in the survival probability with decreasing neutrino energy (Fig.~\ref{fig:survivalProbability}) becomes harder to observe in the electron ratio spectrum, e.g., the $\beta = 30$ and $\beta = 20$ cases are even harder to distinguish than they were in Fig.~\ref{fig:survivalProbability}.  Figure~\ref{fig:predictedSpectrum} (bottom panel) shows that our prediction of the electron spectrum accurately reproduces the Super-K observation, which is an important check.

%%%%%%%%%%%%%%%%%%%%%%%%%%%%%%%%%%%%%
%%%%%%%%%%%%%%%%%%%%%%%%%%%%%%%%%%%%%

%%%%%%%%%%%%%%%%%%%%%%%%%%%%%%%%%%%%%%%%%%%%%%%%%%%%%%%%%%%%%%%%%%%%%%%
%%%%%%%%%%%%%%%%%%%%%%%%%%%%%%%%%%%%%%%%%%%%%%%%%%%%%%%%%%%%%%%%%%%%%%%

\section{Calculation of probes of the solar core}
\label{sec:probes}

In this section, we present our main results, which are the projected uncertainties in the density-temperature plane probed by \B observations, referring to the best-fit values as $\rho_{\textrm{npz}}$ and $T_{\textrm{npz}}$ because neutrinos probe the values in the neutrino production zone.  The density is probed via the neutrino survival probability and the temperature is probed by the flux. We then outline paths toward model-independent neutrino probes of the Sun.

%%%%%%%%%%%%%%%%%%%%%%%%%%%%%%%%%%%%%%%%%%%%%%%%%%%%%%%%%%%%%%%%%%%%%%%

\subsection{Constraints in the \texorpdfstring{$\rho, T$}{rho, T} plane}
\label{subsec:mainResult}

To probe the typical density for \B production, we use $\beta$ to parameterize the shape of the neutrino production zone and hence the neutrino survival probability.  We fit for $\beta$ by optimizing the log-likelihood, 
\begin{equation}
    \mathcal{L} (\beta) = -\frac{1}{2}\sum_i \left[ \left(\frac{O_i -\left(\Delta N/\Delta T\right)_i (\beta)}{\sigma_i} \right)^2  + \ln(2\pi\sigma_i^2)\right],
    \label{eq:testStatistic}
\end{equation}
where $O_i$ is the observed recoil electron spectrum in Super-K~\cite{Super-Kamiokande:2023jbt}, the second term in the quadratic is the predicted spectrum using Eq.~(\ref{eq:ourModel}), and $\sigma_i$ are the statistical uncertainties.

%%%%%%%%%%%%%%%%%%%%%%%%%%%%%%%%%%%%%
\begin{figure}[t]
    \centering
    \includegraphics[width=0.99\linewidth]{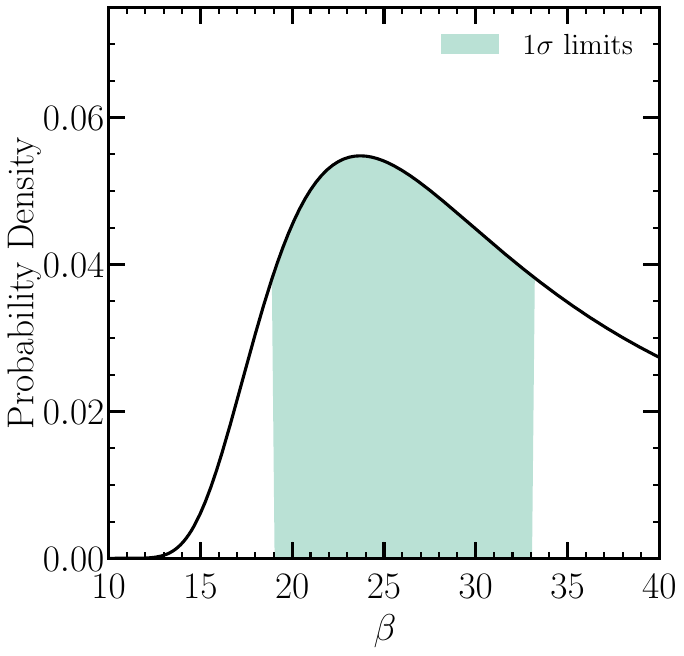}
    \caption{Probability density for $\beta$ given Super-K's observed recoil electron spectrum.  The best fit is \blue{$\beta = 23.7^{+9.2}_{-4.7}$}, where the uncertainties correspond to 68\% of samples (with $34\%$ of samples on each side of the maximum).}
    \label{fig:fitResult}
\end{figure}
%%%%%%%%%%%%%%%%%%%%%%%%%%%%%%%%%%%%%

Figure~\ref{fig:fitResult} shows the resulting probability distribution for $\beta$.  We obtain this by stepping over Eq.~(\ref{eq:testStatistic}) for varying $\beta$ and plotting the Gaussian likelihood (via $\exp({\mathcal{L}(\beta)})$, normalized to unity). For small values of $\beta$, the distribution has a short tail because decreasing $\beta$ corresponds to moving the neutrino production zone to larger radii, thus smaller densities (Fig.~\ref{fig:solarProfiles}), smaller matter angles (Fig.~\ref{fig:matterAngle}), and larger survival probabilities (Fig.~\ref{fig:survivalProbability}), which would conflict with Super-K data (Fig.~\ref{fig:predictedSpectrum}). For large values of $\beta$, the distribution has a long tail because increasing $\beta$, which leads to the opposites of the changes noted above, is less restricted. That is because the survival probability becomes flatter in Super-K's energy range, in part because the solar density profile flattens out at small radii.  Additionally, for smaller $\beta$, the distribution in Fig.~\ref{fig:solarProfiles} becomes narrower, so that the effects of averaging over the neutrino production zone in Eq.~(\ref{eq:source_term}) are less.

\begin{figure}[t]    \centering
    \includegraphics[width=0.99\linewidth]{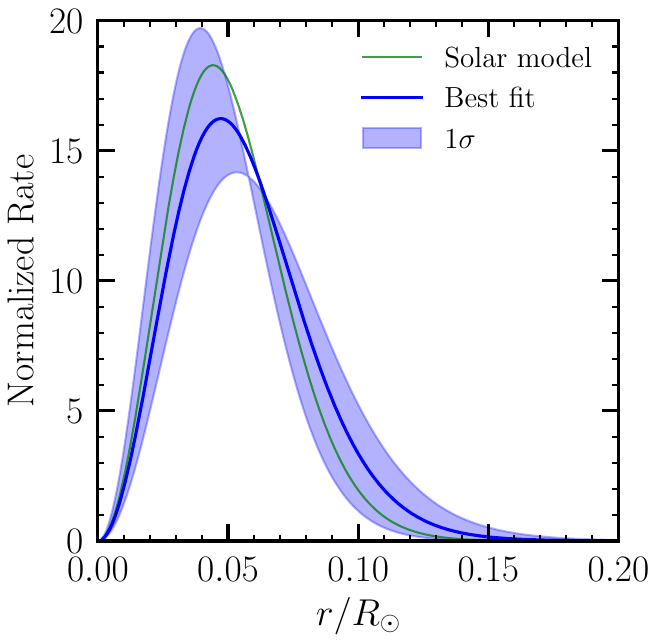}
    \caption{Neutrino production rate profile determined by our fit compared to that from the full calculation using the reference solar model. While our result is shifted relative to the full calculation, the width is similar.}
    \label{fig:fitComparison}
\end{figure}

Figure~\ref{fig:fitComparison} shows the neutrino production rate profile corresponding to the best-fit value of beta.  \blue{We find
$23.7^{+9.2}_{-4.7}$, which is slightly} lower than the value that closely approximates the full calculation using Eqs.~(\ref{eq:Gamma}) and (\ref{eq:sigma_v_approximation}) (where we find $\beta \approx 27$, as shown in Fig.~\ref{fig:NPZ}), but it is adequate for our purposes \blue{of making an initial estimate}. Though the best-fit neutrino production rate profile is somewhat shifted compared to that of the full calculation, the widths of the distributions are comparable.

%%%%%%%%%%%%%%%%%%%%%%%%%%%%%%%%%%%%%
%%%%%%%%%%%%%%%%%%%%%%%%%%%%%%%%%%%%%

To probe the typical temperature for \B production, some assumptions \blue{and approximations} are needed.  As described in Sec.~\ref{subsec:production}, it has been found that the total \B neutrino flux scales as $\phi_\mathrm{tot} \propto T_\textrm{c}^{24}$, where $T_\textrm{c}$ is the core temperature~\cite{2002PhLB..526..186F}.  What we need is a scaling relation in terms of $T_\textrm{npz}$, including the prefactor, determined from contemporary solar models.  This does not seem to exist in the literature, though it should be calculated.  To proceed, we assume the same exponent holds for $T_\textrm{npz}$, which is only 6\% less than $T_\textrm{c}$ in the reference solar model.  That is, we use $\phi_\mathrm{tot} \propto T_\textrm{npz}^{24}$, though the prefactor is unknown.  While this does not allow us to directly extract the \textit{value} of $T_\textrm{c}$ from the SNO measurement of the total \B flux, it does allow us to \blue{estimate the} projected \textit{uncertainty}, which is our goal. 

\magenta{We thus assume $\phi_\mathrm{tot} \propto \phi_{\rm SSM}(T_0) (T_{\rm npz}/T_0)^{24}$, where $\phi_{\rm SSM}(T_0)$ is the predicted standard solar model flux. We consider two cases for the derived uncertainty on $T_{\rm npz}$.  In the first case, we take into account only the uncertainty on the measured flux. Then simple error propagation leads to $\delta\phi_\mathrm{tot}/\phi_\mathrm{tot} = 24~\delta T_\textrm{npz}/T_\textrm{npz}$.  Taking $\delta \phi_\mathrm{tot}/\phi_\mathrm{tot} \approx 3.8\%$ from SNO~\cite{SNO:2011hxd} leads to $\delta T_\textrm{npz}/T_\textrm{npz} \approx 0.2\%$. In the second case, we also take into account the uncertainty on $\phi_{\rm SSM}$ itself, which can be extracted from Table~2 of Ref.~\cite{Haxton:2012wfz}. This combined modeling uncertainty is about 17\%, and is dominated by the atomic opacity ($\approx$$7\%$), the nuclear $S$-factors ($\approx$$5\%$ for $S_{34}$ and $\approx$$ 8\%$ for $S_{17}$), the diffusion coefficient ($\approx$$4\%$), and the Fe abundance ($\approx$$6\%$).  Taking this into account gives an uncertainty on $T_\textrm{npz}$ of $0.7\%$, which which can be improved in the future.}

\begin{figure}[t]
    \centering
    \includegraphics[width=0.99\linewidth]{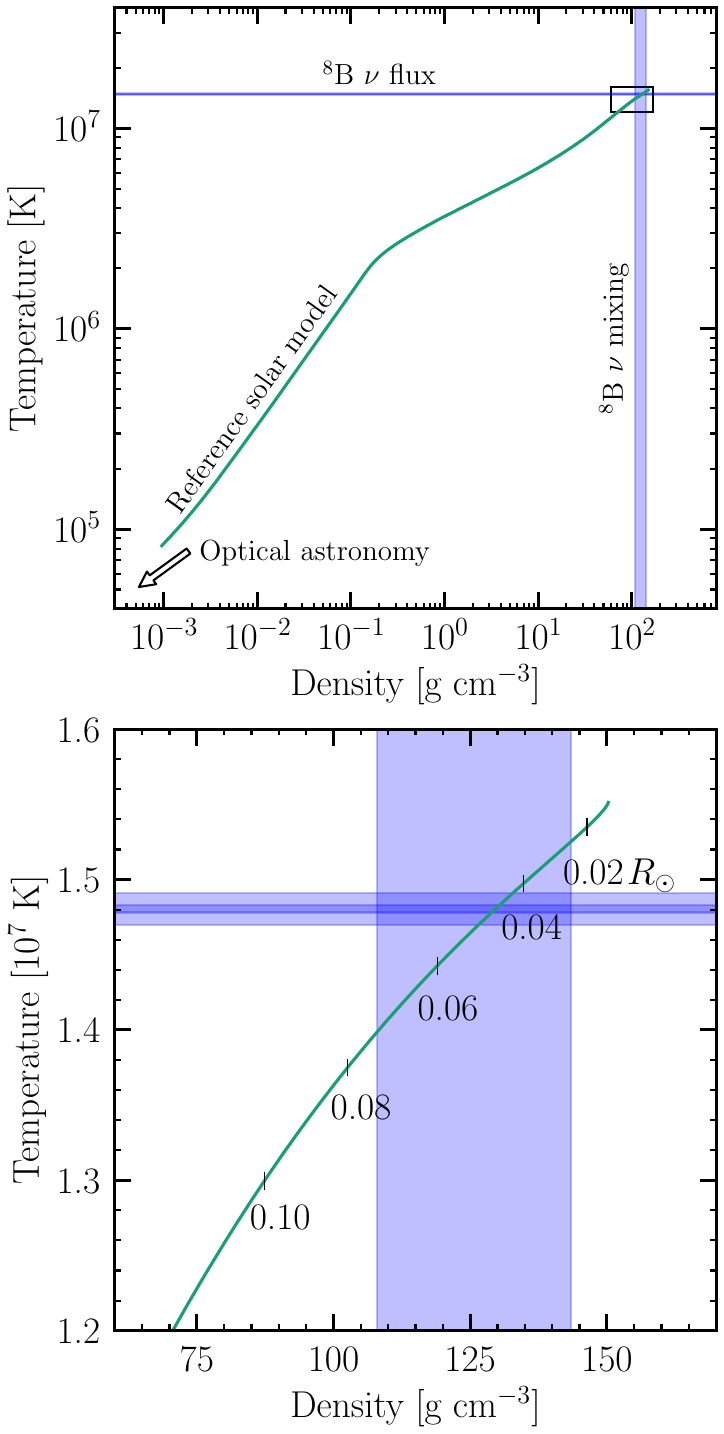}
    \caption{Our \blue{projected estimates of} constraints on the reference solar model density and temperature profiles using \B neutrino observations. \magenta{The thicker horizontal band takes into account modeling uncertainties on the \B flux, and the thinner one only experimental uncertainties, showing the potential impact of reducing modeling uncertainties.}}
    \label{fig:mainResult}
\end{figure}

Figure~\ref{fig:mainResult} shows our main results.  As far as we are aware, \blue{this plot is the first of its kind where joint neutrino probes of the solar interior density and temperature are shown}.  For $\rho_{\textrm{npz}}$ and its uncertainty, we use the peak and width from Fig.~\ref{fig:fitComparison}, where we specify the latter by including 68\% of the distribution, with 34\% on each side of the maximum. \blue{This leads to $\rho_{\textrm{npz}} = 129^{+14}_{-22}$~g~cm$^{-3}$; in the reference solar model, this corresponds to the range of $0.026 \leq r/R_\odot \leq 0.073$}. Note that while we only use the MSW effect and the Super-K electron energy spectrum, we are able to constrain the radial range (and thus angular extent) of the neutrino production zone even without angular information. (Eventually, direct angular measurements may be possible~\cite{Davis:2016hil}.) For $T_{\textrm{npz}}$ and its uncertainty, we use the value from the reference solar model at the peak of the best-fit production profile and the uncertainty estimated above. \magenta{For our optimistic case, }\blue{we find $T_\textrm{npz} = (1.480 \pm 0.002) \times 10^7$~K}. \magenta{Taking into account other uncertainties recounted above, we find $T_\textrm{npz} = (1.480 \pm 0.010) \times 10^7$~K}.

Figure~\ref{fig:mainResult} also shows that neutrinos provide a unique perspective on the Sun, as they probe the deep \textit{interior} of the Sun, which nothing else can do\blue{, which was originally pointed out by Bahcall~\cite{Bahcall:1964gx} and Davis~\cite{Davis:1964hf}}.  Astronomical observations in the optical and other wavebands first probe the \textit{exterior} of the Sun.  With helioseismology, it is possible to probe the solar interior, though not this deep, and less directly, as the primary observable is the  sound speed squared. Figure~\ref{fig:mainResult} (top panel) shows that even though neutrino observations are difficult, they are surprisingly powerful. Figure~\ref{fig:mainResult} (bottom panel) highlights the regions of the solar interior that can be probed. Overall, we find no evidence for significant problems in the physical description of the Sun.

%%%%%%%%%%%%%%%%%%%%%%%%%%%%%%%%%%%%%%%%%%%%%%%%%%%%%%%%%%%%%%%%%%%%%%%

\subsection{Towards broader studies} 
\label{subsec:future}

As discussed in Sec.~\ref{sec:intro}, we are approaching a time when we can probe the astrophysics of the Sun in powerful new ways.  Neutrinos are a unique tool for revealing the physical conditions of the solar interior, but their \blue{particle} properties have been uncertain.  With coming laboratory measurements of neutrino mixing, that uncertainty will be greatly reduced.  In addition, there are excellent prospects for laboratory measurements and computational studies of the basic physics of nuclear reactions and atomic opacities to reduce those input uncertainties. For example, most of the error budget in the theoretical prediction of the total \B flux lies in $S_{17}$. Even so, uncertainties on $S_{17}$ are about $\approx$3\%~\cite{Acharya:2024lke}, but could become even smaller via new efforts in nuclear theory and experiment. These changes mean that we can focus our attention on things that \textit{can only be measured with solar \blue{neutrino} data}. 

A theme illustrated by our work is that a true inversion from neutrino observations to solar-core parameters is not possible, even in the simplified calculations above.  First, there are multiple steps that introduce significant smearing.  Second, and more fundamentally, the Sun is an enormously complicated system with many variables.  It only makes sense to consider changes to the solar core density and temperature that are consistent with the equations of stellar structure and wide variety of observational constraints.  Therefore, one should consider the forward problem, where uncertain physical aspects of solar models are varied and their effects are carried through to neutrino and electromagnetic observables. This kind of approach can help address longstanding mysteries about the Sun.

For instance, there is discrepancy between heavy element abundances in the solar photosphere~\cite{2014dapb.book..245B, 2014ApJ...787...13V, 2020arXiv200406365V} and those in the interior implied from helioseismology~\cite{Basu:2007fp, 2012ApJ...746...16V, Christensen-Dalsgaard:2018etv}. Though much work has been done to study this tension using neutrinos~\cite{1997MNRAS.289L...1A, 2003ApJ...599.1434C, Serenelli:2012zw, 2018MNRAS.477.1397S} and additional approaches~\cite{Song:2017kvf, Kunitomo:2022flu}, a solution to the solar metallicity problem has yet to be found.  Other questions about solar models remain, for example, changing the compositions in the solar interior or sufficiently perturbing the microphysics~\cite{Yang:2024ocp, 2024MNRAS.534.2968B}. Specifically, the effects of including chemical mixing in the radiative zone via rotation or waves in solar models have yet to be explored.

What is needed is for solar modelers to create large suites of models that sample over uncertainties in solar astrophysics.  Decades ago, large suites of solar models were created by varying inputs via Monte Carlo~\cite{Bahcall:1996vj}.  However, given the computational constraints of the time, these models had to be characterized in simple ways, e.g., via scatterplots of the neutrino flux from one nuclear reaction versus another.  It is now possible that the complete models could be shared.  Or the solar modelers could collaborate with neutrino experimentalists and phenomenologists to accurately test which models actually remain allowed.  In this way, the neutrino data would probe solar astrophysics in new ways.

In the future, there will also be better observations to test these phenomena. On the neutrino side, these observables include the fluxes and spectra of neutrinos from several different nuclear reactions~\cite{2016JHEP...03..132B}, which have different sensitivities to changes in the physical conditions of their production zones. For example, better measurements of the \textit{pp} neutrino flux would improve constraints on the Sun's luminosity~\cite{Bahcall:2001pf}. JUNO is poised to make improved measurements of $^7$Be, \textit{hep}, and CNO neutrinos~\cite{2023JCAP...10..022A}, which, in addition to \B~\cite{JUNO:2020hqc, JUNO:2022jkf}, would be helpful in addressing the solar metallicity problem~\cite{2020arXiv200406365V}. Newly precise measurements of \B solar neutrinos can also be made by Hyper-Kamiokande~\cite{Hyper-Kamiokande:2016srs, 2018arXiv180504163H} and DUNE~\cite{Capozzi:2018dat, Parsa:2022mnj, Meighen-Berger:2024xbx}. Measurements of \B neutrinos at as low energies as possible would help observe the upturn in the survival probability (Fig.~\ref{fig:survivalProbability}). To make these \B measurements as precise as possible, it would be important to reduce detector backgrounds even further~\cite{Zhu:2018rwc, Nairat:2024upg}. Studying the conditions of the solar core using neutrinos from \textit{pep} and $^7$Be will require high precision because these lower-energy neutrinos experience weak matter effects (Fig.~\ref{fig:matterAngle}) as they are born at densities below the MSW resonance. The first measurements of the \textit{hep} neutrino flux and spectrum would confirm that the rare pp-IV chain does indeed occur in the Sun and open the door to studying a different radial range inside the Sun around $\approx$$0.15 R_\odot$. On the electromagnetic side, the primary need is for new efforts in helioseismology theory to better understand existing observations~\cite{2013SoPh..287....9G, 2016LRSP...13....2B}.

Our analysis of \B neutrinos could be improved.  With varying solar models as described above, one could make direct calculations of their compatibility with \B data. In this \blue{first} paper, we generate constraints on $\rho_\textrm{npz}$ and $T_\textrm{npz}$ separately. We took this approach because it is both semi-solar-model-independent and simple. By assuming the SNO measurement for $\phi_\mathrm{tot}$ and using $\Gamma(r) \propto T(r)^\beta$, we ignored changes in the \B flux \blue{for the purpose of computing initial estimates}; taking those into account would give greater sensitivity. \blue{In upcoming studies}, it would be interesting to consider the charged-current neutrino-deuteron data from SNO, as the differential cross section is narrower and only electron neutrinos contribute, though the statistics are lower. \blue{Separately, it would be interesting to consider $^{7}$Be and \textit{hep} data and how those would produce additional constraints on the solar temperature and density in their corresponding neutrino production zones.}

In the long term, joint advances in neutrino physics and solar astrophysics will allow more sensitive probes of new physics. Axion emission~\cite{2015JCAP...10..015V}, for example, is very sensitive to the solar temperature profile. Measurements of different solar neutrinos would allow for probing a different range of axion masses and coupling constants. Likewise, neutrino data have recently been used to constrain the mass of dark cores in the Sun~\cite{Bellinger:2025hrg}. Measurements of \B neutrinos, including the flux and survival probability, may be sensitive tools to study such non-standard effects.

%%%%%%%%%%%%%%%%%%%%%%%%%%%%%%%%%%%%%
%%%%%%%%%%%%%%%%%%%%%%%%%%%%%%%%%%%%%

%%%%%%%%%%%%%%%%%%%%%%%%%%%%%%%%%%%%%%%%%%%%%%%%%%%%%%%%%%%%%%%%%%%%%%%
%%%%%%%%%%%%%%%%%%%%%%%%%%%%%%%%%%%%%%%%%%%%%%%%%%%%%%%%%%%%%%%%%%%%%%%

\section{Conclusions and Outlook}
\label{sec:conclusions}

\blue{Following the solution of the solar-neutrino problem two decades ago~\cite{SNO:2001kpb, Super-Kamiokande:2001ljr, KamLAND:2002uet}, this topic has receieved less attention. But we didn't run out of questions --- we ran out of data.} In the next decade, this will change significantly, as JUNO, Hyper-Kamiokande, DUNE, and other experiments (e.g., possibly XLZD~\cite{XLZD:2024nsu}) come online. Those experiments will make much more precise measurements of solar neutrino fluxes and spectra.  And, because these and other experiments will also make much more precise measurements of neutrino mixing parameters through independent measurements of laboratory neutrino sources, we have new opportunities to use solar neutrinos to probe the astrophysics of the solar interior.

Even with present solar-neutrino data, there will be important steps that can be taken once the neutrino mixing parameters are better known. Building on Refs.~\cite{Bahcall:1989ks, Bahcall:1996vj}, we show that the total flux of \B neutrinos measured by SNO \blue{(\textit{without} the effects of active-flavor oscillations)} probes the temperature of the neutrino production zone, $T_{\textrm{npz}}$. Building on Refs.~\cite{1997PhDT.........9B, Balantekin:1997fr, Lopes:2013nfa, Lopes:2013sba, Laber-Smith:2022eih, Laber-Smith:2024hbc}, we show that the oscillated flux of \B neutrinos measured by Super-K \blue{(\textit{with} the effects of active-flavor oscillations)} probes the density of the neutrino production zone, $\rho_{\textrm{npz}}$. \blue{While both our constraints arise from \B data, it is not the same data.} Last, we are the first to combine these to obtain projected joint constraints on the density-temperature plane. Using straightforward treatments of neutrino production, propagation, and detection, we show that we can expect to obtain tight constraints in that plane with only minor reliance on solar models. 

Quoting from Bahcall's 1964 paper~\cite{Bahcall:1964gx}, ``Only neutrinos, with their extremely small interaction cross sections, can enable us to \underline{see into the interior of a star} and thus verify directly the hypothesis of nuclear energy generation in stars."  More than sixty years later, this dream is within reach.  Neutrinos take patience, but they reward it richly.

The Python code used to generate our results and the figures in this paper, as well as the relevant data files, are all available on \href{https://github.com/melanieAzaidel/solarLMAO}{GitHub}.

\bigskip

\textbf{\textit{Note added:}} In the final stages of preparing our paper, there appeared a new paper (\magenta{Ref.~\cite{2025PhLB..86639560D}}) that addresses similar directions but with a different approach and complementary results.

%%%%%%%%%%%%%%%%%%%%%%%%%%%%%%%%%%%%%%%%%%%%%%%%%%%%%%%%%%%%%%%%%%%%%%%
%%%%%%%%%%%%%%%%%%%%%%%%%%%%%%%%%%%%%%%%%%%%%%%%%%%%%%%%%%%%%%%%%%%%%%%

\medskip

\section*{Acknowledgments}

We are grateful for helpful discussions with Peter Denton, Ivan Esteban, Dick Furnstahl, Yifan Jiang, Ming-Wei Li, Obada Nairat, Stephen Parke, Marc Pinsonneault, Georg Raffelt, Michael Smy, Yasuo Takeuchi, \magenta{and the anonymous referee}. M.A.Z.\ acknowledges this material is based upon work supported by the National Science Foundation Graduate Research Fellowship under Grant No.\ DGE-2240614. The work of J.F.B.\ was supported by National Science Foundation Grant No.\ PHY-2310018.

%%%%%%%%%%%%%%%%%%%%%%%%%%%%%%%%%%%%%%%%%%%%%%%%%%%%%%%%%%%%%%%%%%%%%%%
%%%%%%%%%%%%%%%%%%%%%%%%%%%%%%%%%%%%%%%%%%%%%%%%%%%%%%%%%%%%%%%%%%%%%%%

\bibliography{refs}

\end{document}